\documentclass[]{elsarticle}
\usepackage{epstopdf} 
\usepackage{color}
\usepackage{lscape}
\biboptions{sort&compress}
\usepackage{graphicx}
\RequirePackage{amsmath}
\begin{document}
\begin{frontmatter}
\title{On the nature of the glass transition in metallic glasses after deep relaxation }

\author[mymainaddress]{A.S. Makarov}
\author[mymainaddress]{G.V. Afonin}
\author[mymainaddress]{R.A. Konchakov}
\author[mymainaddress2]{J.C. Qiao}
\author[mymainaddress5]{N.P. Kobelev}
\author[mymainaddress]{V.A. Khonik}
\corref{mycorrespondingauthor}
\ead{v.a.khonik@yandex.ru}
\address[mymainaddress]{Department of General Physics, State Pedagogical University, Lenin Street 86, Voronezh 394043, Russia}
\address[mymainaddress2]{School of Mechanics and Civil Architecture, Northwestern Polytechnical University, Xi'an 710072, China}
\address[mymainaddress5]{Institute of Solid State Physics, Russian Academy of Sciences, Chernogolovka, Moscow district 142432, Russia}

\begin{abstract}
We performed parallel study of calorimetric and high-frequency shear modulus behavior of Zr-based metallic glasses after deep relaxation just below the glass transition. It is shown that deep relaxation results in the appearance of a strong  peak of the excess heat capacity while the shear modulus is moderately affected. A theory assuming high-frequency shear modulus to be a major physical parameter controlling glass relaxation is suggested. The energy barrier for these rearrangements is proportional to the shear modulus while its magnitude, in turn, varies due to the changes in the defect concentration (diaelastic effect). Both dependences lead to the occurrence of  heat effects. The excess heat capacity calculated using experimental shear modulus data demonstrates very good agreement with the experimental calorimetric  data for all states of glasses. It is argued that the glass transition behavior after deep relaxation of glass is close to a phase transition of the first kind.
\end{abstract}

\end{frontmatter}

\section{Introduction}

The nature of the glass transition  still remains one of the most controversial issues in the physics of glasses \cite{NelsonPRB1983,Nemilov1995,DyreRevModPhys2006,Varshneya2019} and this is in full applied to metallic glasses (MGs)  \cite{WeiNatCommun2013,AlbertScience2016,SanditovPhysUsp2019,DuMaterToday2020,DongAPL2021,ShenActaMater2022,KonchakovJNCS2023}.
In principle, there are currently two main approaches to this problem. The first approach considers the glass transition as a purely kinetic process \cite{DyreRevModPhys2006,SanditovPhysUsp2019} while the second one treats  it as a phase transition of the second kind with some features related to the fact that glass is not an  equilibrium system \cite{IUPAC1997,AlbertScience2016,Schmelzer2011,KonchakovJNCS2023}. One of the ways to derive valuable information on this problem is related to detailed in-depth studies of heat capacity  behavior at temperatures close to the glass transition temperature $T_g$, both experimentally and theoretically. 

The number of such studies for MGs is quite limited. One of the main reasons for this is due to the fact that researchers usually study as-cast or weakly relaxed MGs and in this case relatively small changes in the heat capacity are observed near the glass transition.  This is in contrast to non-metallic glasses, where the presence of heat capacity peaks near the glass transition  has been well documented. In fact, even in metallic glasses, such peaks can be observed after deep relaxation \cite{ChenJNCS1981,BuschApplPhysLett1998,HaruyamaActaMater2010,EvensonActaMat2011,HaruyamaMaterTrans2014,AfoninPSSRRL2018}.
Since the presence of heat capacity peaks can be considered as a characteristic sign of either a phase transition of the second kind or another critical phenomenon, further investigation of this phenomenon is highly relevant. Therefore, the purpose of this study is to investigate the behavior of the heat capacity in MGs with different levels of relaxation, including deep relaxation. Our investigation has led to a rather unexpected conclusion: the glass transition of deeply relaxed MGs exhibits features similar to those of a phase transition of the first kind.

\section{Relationship between glass heat capacity and changes of the shear modulus }

The properties of glass are intrinsically related to an important physical quantity -- the instantaneous shear modulus because it \textit{i}) controls the heights of barriers for local atomic rearrangements \cite{Nemilov1995,DyreRevModPhys2006} and \textit{ii}) constitutes a thermodynamic parameter, as it is the second derivative of the Gibbs free energy with respect to shear strain \cite{Hirth,GranatoPRL1992}. Meanwhile, the instantaneous   shear modulus (in practice, high frequency shear modulus called simply shear modulus hereafter) is a major ingredient of the Interstitialcy theory (IT), which was originally suggested by Granato  \cite{GranatoPRL1992,GranatoEurJPhys2014} and further developed in numerous works (a review is  given in Ref.\cite{KobelevUFN2023}). In our viewpoint, the IT currently constitutes one of the most successful approaches to the understanding of MGs' relaxation behavior (a recent example is given in Ref.\cite{KhmyrovMatChemPhys2025}). The IT is based on the statement that  the melting of metals is related to an avalanche-like increase in the number of interstitials in the dumbbell form, which destabilize the crystalline lattice and lead to a sharp drop in the shear modulus, as  confirmed by indirect experimental observations \cite{SafonovaJPCM2016,SafonovaJETPLett2016,GoncharovaJETPLett2017,KobelevUFN2023}. In the liquid state, these entities remain identifiable structural units while melt quenching freezes them in the solid glass. 

The structure of the glass becomes significantly heterogeneous. Most of interstitial-type defects (otherwise understood as elastic dipoles \cite{KobelevJApplPhys2014}) gather in clusters formed by 5-7 defects with dominating icosahedral  symmetry \cite{KobelevUFN2023,KonchakovJPhysConMatt2019}. They represent an amorphous matrix, which is quite stable until the beginning of crystallization. A significantly smaller part of the structure consists of single defects or small clusters of 2-3 defects, which are more mobile and less stable. It is these defects that define the relaxation behavior of MGs, and they are discussed below.

The estimates show that the concentration of dumbbell interstitials required for melting and, accordingly, for  glass formation  is 5--10\%. At the same time, the concentration of defects involved in relaxation processes in glass lies in the range of 0.2--0.5\%, i.e. it is less than one tenth of the total concentration of dumbbell interstitials assumed to participate in melting \cite{KobelevUFN2023}. According to the IT, the shear modulus of glass exponentially depends on the concentration of interstitial-type defects. Temperature results in thermoactivated fluctuation atomic rearrangements, which can be understood as defect annihilation/creation and/or alteration of defect energy states \cite{KobelevJALCOM2021}. All this constitutes structural relaxation. 

Elementary events of structural relaxation require thermal activation, that is doing the work $r$ against the elastic resistance force of the medium, which is determined by the instantaneous shear modulus $G$. i.e.

 \begin{equation}
 r=GV_0, \label{r}
 \end{equation}
where $V_0$ is a characteristic volume of atomic rearrangements in the vicinity of  a defect. Then, for the implementation of $N$ structural rearrangements per mole one should do the work

\begin{equation}
R=Nr=NGV_0. \label{R}
\end{equation}
Therefore, for the elementary work $\delta R$ one can write down 
\begin{equation}
 \delta R=d(G NV_0)=V_0\left(GdN+NdG\right). \label{deltaR} 
 \end{equation} 
 
Let us  consider the first term of this formula associated with the production or disappearance of the aforementioned defects. According to the IT, the change in the concentration $c$ of interstitial-type defects ($c=N/N_A$, where $N_A$ is the Avogadro number) is related with the  change of the instantaneous shear modulus (diaelastic effect) as

\begin{equation}
dc=-dG/B G,  \label{delta c}
\end{equation}
where $B$ is a dimensionless so-called shear susceptibility. Then, for the elementary work one finds
\begin{equation}
\delta R=N_AV_0dG/B.   \label{delta R}
\end{equation}

In the case of an arbitrary number of fluctuations in the entire volume of the body, it is necessary to spend the work $R_d=\int \delta R$, where $\delta R$ is the elementary work upon fluctuations occurring in an infinitely small volume $dV$.  If one knows $R_d$, it is possible to calculate the entropy change $\Delta S_d$ of the whole body due to fluctuations in its small part and corresponding heat effect can then be accepted as
\begin{equation}
\delta Q=Td\Delta S_d=-\delta R_d=-\frac{N_AV_0}{B}d\Delta G_d,    \label{delta Q}
\end{equation}
where $\Delta G_d$ is the change of the shear modulus related to fluctuations. 

Therefore,  the heat capacity of the body $\Delta C_d$ related to thermal fluctuation becomes
\begin{equation}
 \Delta C_d=T\frac{d \Delta S_d}{dT}=-\frac{N_AV_0}{B}\frac{d\Delta G_d}{dT}. \label{Delta Cd}
 \end{equation} 

Thus, defect production or disappearance leads to changes of the shear modulus and provides a contribution to the heat capacity given by Eq.(\ref{Delta Cd}). It was earlier demonstrated that the heat capacity determined by this formula is  satisfactorily consistent with the temperature dependence of the heat capacity calculated from calorimetry data \cite{KonchakovJNCS2023}. 

Let us now consider the second term in Eq.(\ref{deltaR}), which can be written as

\begin{equation}
\delta R_f=N_AV_0c_rdG,    \label{dRf}
\end{equation}
where $c_r$ is the concentration of defects taking part in the relaxation process. 
It is seen that this term describes the change in the elementary work in the volume of the body associated with fluctuations upon changes of shear modulus.   According to fluctuation thermodynamics, the work $R_f$ characterizes a quasi-equilibrium change in the Gibbs free energy 
$\Phi_f$ of the defects involved in fluctuation rearrangements  describing the transitions of the defect subsystem from one quasi-equilibrium energy state to another. This leads to a change of the  entropy given as   
\begin{equation}
\Delta S_f=-\frac{d\Delta \Phi_f}{dT}=-N_AV_0c_rdG/dT.   \label{Delta Sf}
\end{equation}
Accordingly, there  appears an additional contribution to the heat capacity equal to
\begin{equation}
\Delta C_f=T\frac{d\Delta S_f}{dT}=-N_AV_0\left[c_rT\frac{d^2G}{dT^2}+T\frac{dc_r}{dT}\frac{dG}{dT}\right].   \label{Cf}
\end{equation}

Thus, using Eqs. (\ref{Delta Cd}) and (\ref{Cf}) one can conclude that the total excess heat capacity associated with the defect system of a metallic glass can be accepted as
\begin{equation}
\Delta C_G=\Delta C_d+\Delta C_f=-\frac{N_AV_0}{B}\left[\frac{d\Delta G_d}{dT}+c_rBT\frac{d^2G}{dT^2}+BT\frac{dc_r}{dT}\frac{dG}{dT}\right]. \label{Cp}
\end{equation}

The defect-induced contribution to the shear modulus $\Delta G_d$ in Eq.(\ref{Cp}) can be determined from  temperature dependences of the shear modulus $G$ at temperatures below $T_g$ \cite{MakarovJPCM2022}. However, there is another way to derive this quantity using the IT framework. For this, one can use the basic equation of the IT, which states that  $G=\mu\; exp(-Bc)$, where $\mu$ is the shear modulus of the maternal crystal, $c$ is the total defect concentration and $B\approx 20$ is the shear susceptibility  \cite{KobelevUFN2023}. Using this equation, instead of Eq.(\ref{delta c}) one obtains $\frac{d\Delta G_d}{dT}=-G \frac {dln (\mu/G)}{dT}$. Besides that, using the same equation one can obtain $c_r=ln(\frac{\mu}{G}\frac{G_m}{\mu_m})/B$, where the subscript  $m$ denotes the values of the moduli at a minimum $c_r$-value (this is significant for the as-cast MGs' state; for relaxed states the subscript $m$ denotes the room temperature). As a result, Eq.(\ref{Cp}) can be reduced to the form  
\begin{equation}
\Delta C_G=\frac{N_AV_0}{B}\left[G\frac{dln\frac{\mu}{G}}{dT}\left(1-T\frac{dlnG}{dT}\right)-T ln\left(\frac{\mu}{G}\frac{G_m}{\mu_m}\right)\frac{d^2G}{dT^2}        \right], \label{Cp fin}
\end{equation}
where $V_0=\gamma \Omega$ with $\Omega$ being the volume per atom and $\gamma\approx 1$.

As can be seen from the comparison of Eqs (\ref{Cp}) and (\ref{Cp fin}), the third term in Eq.(\ref{Cp}) actually leads to some renormalization of the first term. Therefore, measurements of the shear modulus make it possible to separate two contributions to the heat capacity, which have significantly different temperature dependences \cite{KonchakovJNCS2023}. Moreover, it follows from Eqs (\ref{Cp}) and (\ref{Cp fin}) that  the contribution proportional to the second derivative of the shear modulus over temperature  in Ref.\cite{KonchakovJNCS2023} was overestimated (since a coefficient proportional to the defect concentration was omitted).

Thus, an additional task in this work was to check the adequacy of the expressions (\ref{Cp}) and (\ref{Cp fin}) in MGs with different degrees of relaxation. At the same time, it should be specially emphasized that the expressions obtained above are valid for atomic rearrangements via thermal fluctuation and are not supposed to describe the processes occurring during crystallization, except, perhaps, its initial stages.

\section{Experimental}

Glassy Zr$_{57}$Nb$_5$Al$_{10}$Cu$_{15.4}$Ni$_{12.6}$ (at.\%, Vit106) was chosen as the main object of the study. Some measurements were also performed on a high-entropy glass Zr$_{35}$Hf$_{13}$Al$_{11}$Ag$_8$Ni$_8$Cu$_{25}$. Both glasses were obtained by quenching  into a copper mold and X-ray verified to be fully amorphous.

Relaxed states were obtained by heating the samples at a rate of 3 K/min to a prescribed  temperature (653 K for Vit106 and 700 K for Zr$_{35}$Hf$_{13}$Al$_{11}$Ag$_8$Ni$_8$Cu$_{25}$), holding at this temperature for certain time and cooling to room temperature at about the same rate. Heat treatments were carried out in evacuated quartz vials. Samples for different types of measurements (calorimetry/shear modulus) were annealed simultaneously in the same vial.

Differential scanning calorimetry (DSC) was performed with a Hitachi DSC 7020 instrument in  high-purity (99.999\%) nitrogen atmosphere using 50-70 mg samples. A crystallized sample of the same composition and nearly the same mass was placed in the reference cell, so that the
instrument measured the difference in the heat flow between the glassy and crystalline samples. This difference is referred to as the differential heat flow $\Delta W$ hereafter.

The electromagnetic acoustic transformation (EMAT) method \cite{VasilievUFN1983} was used to measure the transverse resonant frequencies $f$ (500-700 kHz) of samples ($5\times 5\times 2$ mm$^3$) in a vacuum of about 0.01 Pa. Frequency scanning was  performed automatically  during  heating every 10-15 seconds. The shear modulus $G$ (proportional to $f^2$) as a function of temperature was calculated as its value at room temperature multiplied by the square of the relative change in the resonant frequency. The error in measuring the temperature dependence of $G$ was about 5 ppm near room temperature and about 100 ppm near $T_g$.  Some measurements of the shear modulus at room temperature as well as upon heating were performed by resonant ultrasonic spectroscopy (RUS) at comparable frequencies  using a setup similar to that described in Ref.\cite{BalakirevRevSciInstrum2019} with $3\times 3\times 3$ mm$^3$ samples. The data treatment procedure was the same as that used for EMAT data. The results obtained by EMAT and RUS techniques were very close.

\section{Results}

\begin{figure}[t]
\centering
\includegraphics[scale=0.7]{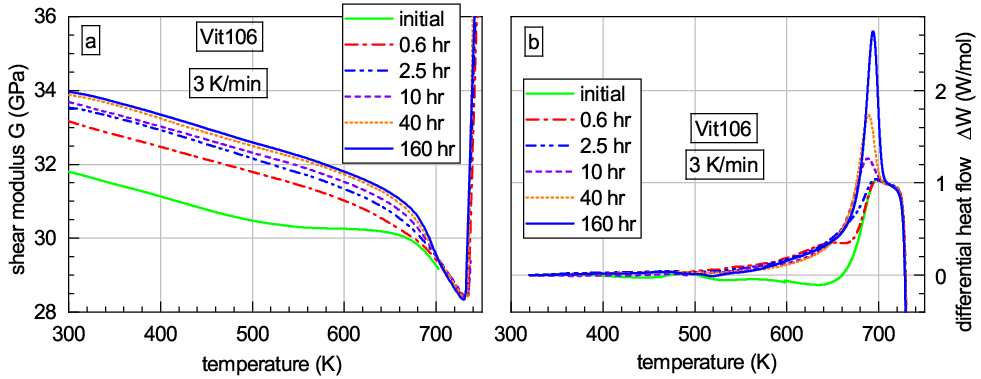}
\caption[*]{\label{Fig1.eps} Temperature dependences of the shear modulus (a) and differential heat flow (b) for Vit106 glass in the initial and relaxed states obtained by preannealing at $T=653$ K during indicated times.}
\label{Fig1}
\end{figure}

Figure \ref{Fig1.eps} shows temperature dependences of the shear modulus $G$ (a) and differential heat flow $\Delta W$ (b) of Vit106  in the initial state and after different preannealing treatments as indicated. It is seen that temperature dependences of $G$ are quite typical. As always, temperature dependence of $G$ changes significantly as a result of the first relaxation (i.e. after 0.6 hour preannealing) while there are no apparent qualitative differences in the behavior of $G$ when the preannealing time is increased, except for its gradual growth at room temperature. At the same time, temperature dependence of the differential heat flow  changes significantly with the degree of relaxation: a pronounced $\Delta W$-peak arises near the glass transition and its height increases with the preannealing time.

It is reasonable to calculate the excess entropy $\Delta S$ of solid glass with respect to its maternal crystalline state. In line with general thermodynamic definition of the entropy, this quantity can be determined as \cite{MakarovJPCM2021}

\begin{equation}
\Delta S(T)=\frac{1}{\dot{T}}\int_{T}^{T_{cr}} \frac{\Delta W(T)}{T}dT, \label{DeltaS}
\end{equation}
where $T$ is the current temperature,  $\Delta W$ is the differential heat flow specified above, $T_{cr}$ is the temperature of the complete crystallization and $\dot{T}$ is the heating rate. This equation shows that  if   current temperature $T=T_{cr}$ then the integral (\ref{DeltaS}) turns to zero and, therefore, $\Delta S$ describes solely the excess entropy of glass with respect to the maternal crystalline state. 

\begin{figure}[t]
\centering
\includegraphics[scale=0.7]{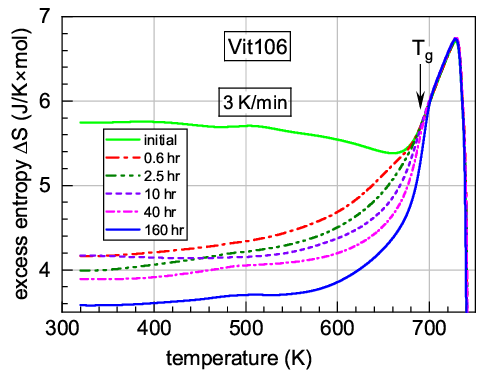}
\caption[*]{\label{Fig2.eps} Temperature dependences of the excess entropy $\Delta S$ calculated with Eq.(\ref{DeltaS}) using  calorimetric $\Delta W$-data shown in Fig.\ref{Fig1.eps}(b) for Vit106 glass preannealed during indicated times. The glass transition temperature $T_g$ is shown by the arrow. It is seen that the derivative  $\frac{d\Delta S}{dT}$  is the largest just below $T_g$ for the longest pre-annealing time of 160 hours.} 
\label{Fig2}
\end{figure}

Temperature dependences of $\Delta S$ calculated with Eq.(\ref{DeltaS})  using  $\Delta W$-data for different preannealing times shown in Fig.\ref{Fig1.eps}(b) for Vit106 glass are given in Fig.\ref{Fig2.eps}. It is seen, first, that preannealing strongly reduces $\Delta S$ below 550--600 K indicating relaxation-induced growth of structural order, just as one would expect. Second, all $\Delta S(T)$-curves merge near $T\approx 690$ K reflecting the transition to the supercooled liquid state, in which the memory of the thermal prehistory is lost. Thus, the temperature of $\approx $690 K constitutes the glass transition temperature $T_g$. The rapid increase of $\Delta S$ with temperature at $T> T_g$ indicates the fast rise of  structural disorder in the supercooled liquid state, which is independent of the preannealing time. The fall of $\Delta S$ to zero above 740 K is due to the complete crystallization.  Finally, it should be emphasized that the excess entropy $\Delta S$ increases most rapidly with temperature just below $T_g$ after the longest preannealing time of 160 hours. This seems to be a very important observation as discussed below.       

\section{Discussion}

\subsection{Relation between   calorimetric heat capacity and heat capacity derived from shear modulus relaxation}

\begin{figure}[t]
\centering
\includegraphics[scale=0.7]{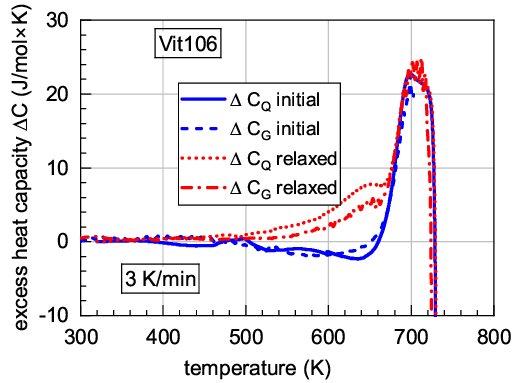}
\caption[*]{\label{Fig3.eps} Temperature dependences of the calorimetric excess heat capacity $\Delta C_Q$ and excess heat capacity $\Delta C_G$ calculated from measurements of the shear modulus using Eq.(\ref{Cp fin}) with $B=20$ and $\gamma=0.6$ for Vit106 glass in the initial state and after relaxation performed by preannealing at $T=653$ K for 0.6 hr.}
\label{Fig3}
\end{figure}

\begin{figure}[t]
\centering
\includegraphics[scale=0.7]{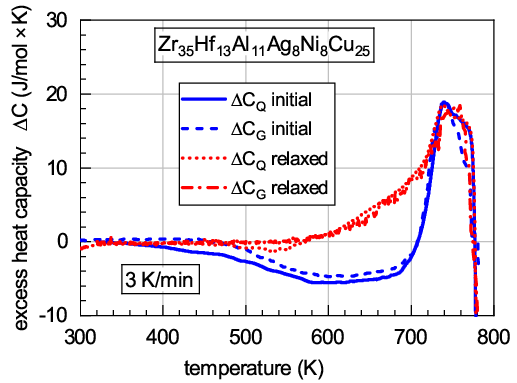}
\caption[*]{\label{Fig4.eps} Temperature dependences of the calorimetric excess heat capacity $\Delta C_Q$ and excess heat capacity $\Delta C_G$ calculated from measurements of the shear modulus using Eq.(\ref{Cp fin}) with $B=20$ and $\gamma=1$ for Zr$_{35}$Hf$_{13}$Al$_{11}$Ag$_8$Ni$_8$Cu$_{25}$ glass in the initial state and after relaxation obtained by preannealing at $T=700$ K for 0.6 hr.}
\label{Fig4}
\end{figure}

Let us  analyze the relationship between temperature dependences of the excess heat capacities obtained directly from measurements of the heat flow ($\Delta C_Q=\Delta W/\dot{T}$) and calculated with Eq.(\ref{Cp fin}) using shear modulus data ($\Delta C_G$). Figure \ref{Fig3.eps} shows temperature dependences of $\Delta C_Q$ and $\Delta C_G$ in Vit106 in the initial and weakly relaxed (0.6 hr preannealing) states. It is seen that calculated temperature dependences $\Delta C_G(T)$ quite accurately describe the experimental $\Delta C_Q(T)$-data. In particular, this applies to the transition from exothermal reaction  in the initial glass below $T_g$ to endothermal reaction upon approaching the glass transition. In addition, the calculated values of the excess heat capacity in the glass transition range also coincide with the experiment. Similar relationship between the calculated and experimental dependences of the excess heat capacity is observed in  high-entropy glassy Zr$_{35}$Hf$_{13}$Al$_{11}$Ag$_8$Ni$_8$Cu$_{25}$ as demonstrated in Fig.\ref{Fig4.eps}. As it is seen in this case, the calculated dependence $\Delta C_G(T)$ also adequately reflects the transition from exothermal to endothermal behavior as a result of relaxation below $T_g$ and is consistent with the experiment in terms of the absolute values of the excess heat capacity. 

\begin{figure}[t]
\centering
\includegraphics[scale=0.7]{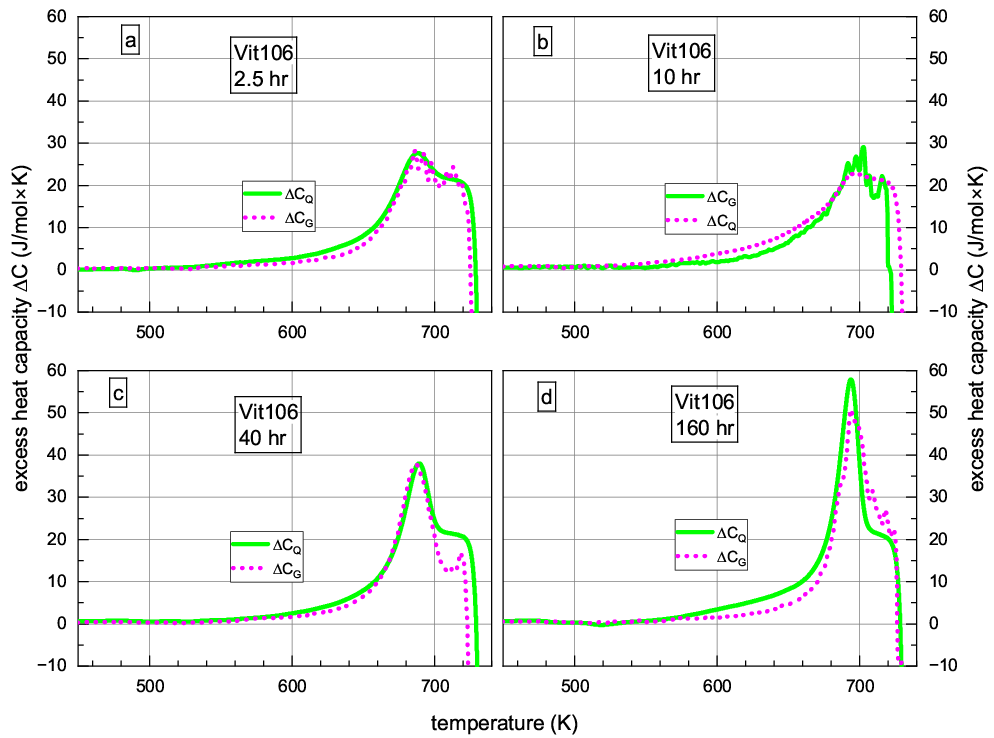}
\caption[*]{\label{Fig5.eps} Temperature dependences of the calorimetric excess heat capacity $\Delta C_Q$ and the excess heat capacity $\Delta C_G$ calculated from shear modulus data using Eq.(\ref{Cp fin}) with $B=20$ and $\gamma=0.6$ for Vit106 samples  preannealed at $T=653$ K for 2.5 hr (a), 10 hr (b), 40 hr (c) and 160 hr (d). It is seen that the calculation provides a good reproduction of calorimetric $\Delta C_Q$-data in all cases. }
\label{Fig5}
\end{figure}

The effect of deep relaxation was studied in detail on Vit106 glass. Figure \ref{Fig5.eps} shows temperature dependences of $\Delta C_Q$ and $\Delta C_G$ for the preannealing times of 2.5 hr, 10 hr, 40 hr and 160 hr. It is seen that $\Delta C_G$-calculation   with Eq.(\ref{Cp fin}) provides a good fit to experimental calorimetric data given by $\Delta C_Q(T)$-dependences in all cases. Thus, it can be concluded that Eq.(\ref{Cp fin}), which is obtained within the framework of the IT, quite adequately describes temperature behavior of the excess heat capacity. Let us discuss  how an analysis of Eqs (\ref{Cp}) and (\ref{Cp fin}) can help in physical  interpretation of the excess heat capacity near the glass transition.

\begin{figure}[t]
\centering
\includegraphics[scale=0.7]{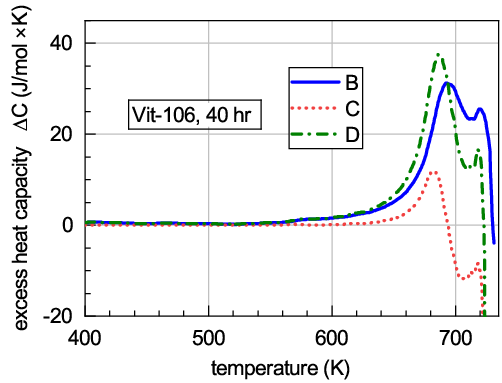}
\caption[*]{\label{Fig6.eps} Temperature dependences of the first (B) and second (C) contributions to the heat capacity according to Eq.(\ref{Cp fin}) together with their sum (D).}
\label{Fig6}
\end{figure}  

First, consider how different contributions to the calculated excess heat capacity of a metallic glass relate to each other. Figure 6 shows temperature dependences of the first (left) and second (right) summands in the right-hand side of Eq.(\ref{Cp fin}) for Vit106 after relaxation for 40 hr at 653 K. It is seen that  the peak excess heat capacity is determined mainly by the first summand, which is proportional to temperature derivative of the concentration of defects. The second term determines the rate of the decrease of the excess  heat capacity with temperature after its maximum. As it follows from the derivation of Eq.(\ref{Cp fin}), the second summand term is proportional to the second derivative of the shear modulus over temperature. At the same time, in the $T_g$-region, the shear modulus is largely determined by the contribution of defects (diaelastic effect), i.e. its second derivative over temperature should be proportional to the second derivative of the defect concentration. Above $T_g$ (i.e. in the supercooled liquid state), the defect concentration does not depend on the preannealing while this concentration in highly relaxed glass sharply increases with temperature. This means that when passing through $T_g$, the second derivative must at some point turn to zero, and change its sign above $T_g$. With an increase of relaxation degree, the concentration of defects involved into the relaxation process, in accordance with Eq.(\ref{Cp fin}), should increase. Consequently, the second contribution to the heat capacity should increase as well and this rises the asymmetry of the peak of the excess heat capacity (its high-temperature part becomes sharper), just as observed in the experiment (see Fig.\ref{Fig5.eps} and Refs \cite{HaruyamaActaMater2010,HaruyamaMaterTrans2014}).

Let us now consider the interpretation of  the excess heat capacity peak observed in highly relaxed state in the $T_g$-region. Generally speaking, it is necessary to distinguish between the glass transition occurring upon cooling (i.e. the transition from a supercooled  state to the  glassy state) and the reverse process taking place upon transition from a glassy state to the supercooled liquid state upon heating. We are interested in the latter process since it  leads to a peak in the excess heat capacity considered above. It is important to reveal the differences of this transition (glass $\rightarrow$ supercooled liquid) for a metallic glass in the initial (as-cast) and highly relaxed states. 

The structure of the glass in the initial state is not very different from the structure of the supercooled liquid state. In both cases, this structure includes certain number of large clusters forming amorphous matrix and defects representing single elastic dipoles as well as  small clusters formed by 2-3 interstitial-type defects (elastic dipoles). The only difference is that the defects are quite mobile above $T_g$, while they are frozen below $T_g$. Moreover, their number in both states is almost the same. In this case, the process of transition from a solid glassy state to the superccoled liquid is a purely kinetic process of defect "defrosting" and the reverse transition (supercooled liquid state $\rightarrow$ solid glass) constitutes defect "freezing".

A different situation is observed in the case of deeply relaxed glass. During relaxation, the number of defects decreases due to their clusterization and/or embedding into larger clusters. In the extreme case, only vanishing amount of  defects do not disappear that should lead to the formation of a hypothetical "ideal" glassy structure.  In a real situation, this is not achievable. However, in any case, the structure formed during deep relaxation is significantly different from the structure of a supercooled liquid. Rather, it can be considered not as a frozen liquid, but as a non-crystalline solid state albeit rather defective. Therefore, one can expect that the transition to a supercooled liquid state in this case should be similar to a phase transition solid state $\rightarrow$ liquid state.

\subsection{On the origin of the glass transition in deeply relaxed state }

\begin{figure}[t]
\centering
\includegraphics[scale=0.7]{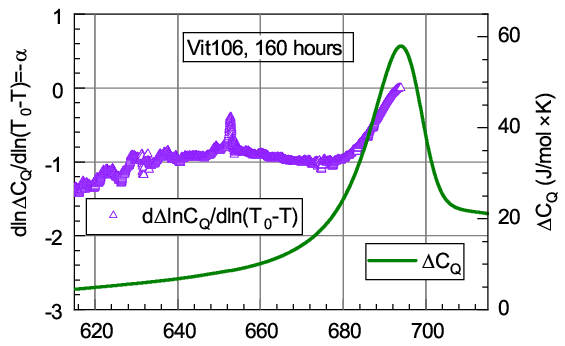}
\caption[*]{\label{Fig7.eps} Temperature dependences of the derivative $\frac{d ln\Delta C_Q}{d ln (T_0-T)}=-\alpha$ and calorimetric excess heat capacity $\Delta C_Q$ for glassy Vit106 relaxed for 160 hours. No indications of either a second kind phase transition ($\alpha\approx 0.1$) or a tricritical point  ($\alpha\approx 0.5$) are observed.}
\label{Fig7}
\end{figure}

\begin{figure}[t]
\centering
\includegraphics[scale=0.7]{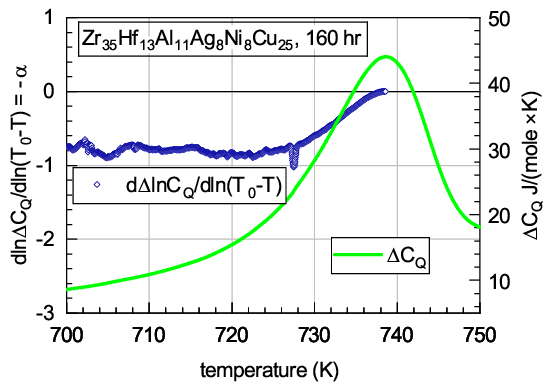}
\caption[*]{\label{Fig8.eps} Temperature dependences of the derivative $\frac{d ln\Delta C_Q}{d ln (T_0-T)}=-\alpha$ and calorimetric excess heat capacity $\Delta C_Q$ for high-entropy glassy Zr$_{35}$Hf$_{13}$Al$_{11}$Ag$_8$Ni$_8$Cu$_{25}$ relaxed for 160 hours.}
\label{Fig8}
\end{figure}

The question  arises what kind of phase transformation this transition represents. Usually, if glass transition is considered as a phase transition, it is assumed to be a phase transition of the second kind (see Ref.\cite{KonchakovJNCS2023} for a review). If this is the case, then upon  approaching the critical temperature of this transition $T_0$ (corresponding to a peak of the excess heat capacity $\Delta C$) a well-defined temperature dependence of $\Delta C$ should be obeyed, i.e. $\Delta C(T) \sim (T_0-T)^{-\alpha}$, where $\alpha\approx 0.1$ for a usual phase transition of the second kind and $\alpha \approx 0.5$ for a tricritical point \cite{StishovJETP2020}. The estimates of this parameter for a few MGs in the initial state give some intermediate value, $0.1<\alpha <0.5$ \cite{KonchakovJNCS2023}. For strongly relaxed MGs in the present investigation, this parameter can be calculated from the relation $\frac{d ln\Delta C_Q}{d ln (T_0-T)}=-\alpha$. Figure \ref{Fig7} gives temperature dependence of this ratio for Vit106 glass after relaxation at 653 K  during 160 hours. It is seen that at temperatures by about  20 K below the glass transition the value of $\alpha$ is close to unity, and when approaching $T_0$ it monotonously decreases  to zero. Thus, there are no indications of any  noticeable temperature range, which can be characterized by either  $\alpha = 0.1$ or $\alpha =0.5$ or by any intermediate $\alpha$-value. Figure \ref{Fig8.eps} shows similar dependence for  high-entropy glassy Zr$_{35}$Hf$_{13}$Al$_{11}$Ag$_8$Ni$_8$Cu$_{25}$, which also underwent deep relaxation by annealing for 160 hours at 700 K. In this case, a similar pattern is observed, i.e. at temperatures noticeably lower than the glass transition, the value of $\alpha$ is close to unity but monotonously tends to zero upon approaching the glass transition.

Thus, the data obtained do not confirm that the  transition of a deeply relaxed glass into a supercooled liquid state is either a phase transition of the second kind or a tricritical point, as it is the case of as-cast MGs \cite{KonchakovJNCS2023}. Besides that, another important fact has to be mentioned. It is generally known that a phase transition of the second kind is not accompanied by the absorption/release of the latent heat. However, the estimates of the heat absorbed by deeply relaxed glass during the transition to a supercooled liquid state give the values of about 1.8 kJ/mole for Vit106 and 1.6 kJ/mole for Zr$_{35}$Hf$_{13}$Al$_{11}$Ag$_8$Ni$_8$Cu$_{25}$. These values are quite significant amounting to 25-30\% of the heat of crystallization (which is close to the heat of melting) of Zr-based MGs \cite{AfoninScrMater2019}. Besides that it is seen from Fig.\ref{Fig5.eps} that the excess heat capacity $\Delta C$ at a given temperature just below the glass transition rapidly increases with the degree of relaxation the (increase of the preannealing time). This is directly manifested in a sharp rise of the excess entropy $\Delta S$ of deeply relaxed glass upon approaching the glass transition while the rate of this rise (i.e. the derivative $\frac{d\Delta S}{dT}$) rapidly increases with the preannealing time  as shown in Fig.\ref{Fig2.eps}. One can then reasonably assume that hypothetical infinitely deep relaxation would result in a discontinuous jump-like increase of the excess  entropy at the glass transition. Meanwhile, such behavior constitutes a major characteristic of the phase transition of the first kind.   

Therefore, a question arises whether the transformation of deeply relaxed glass into a supercooled liquid can be a phase transition of the first kind, i.e. is similar to the melting. Indeed, it is well known that MGs in the supercooled liquid state loose the memory of the preceding thermal prehistory (e.g. see Fig.\ref{Fig2.eps}). Then, the transition of a deeply relaxed glass into the supercooled liquid state should constitute a transformation  from the defect structure consisting of  relatively small amount of low-mobile defects to the structure containing a large number of highly mobile defects characteristic of the liquid. This requires heat supply and it is this endothermal process, which is manifested in the occurrence  of a large heat capacity peak discussed above. 

 The situation becomes similar to the melting of a crystal from the viewpoint of the Interstitialcy theory \cite{GranatoPRL1992,KobelevUFN2023}, which provides a good description of the heat capacity near the glass transition as shown in Figs \ref{Fig3.eps} to \ref{Fig5.eps}: in order for melting to occur, a large number of highly mobile defects such as dumbbell interstitials (elastic dipoles) must be generated. Thus, it is highly likely  that the deeper the relaxation of a metallic glass, the closer the transition glass $\rightarrow$ supercooled liquid will be to the melting, i.e. to a phase transition of the first kind. This constitutes the main idea of the present work. For further verification of this idea, additional studies on metallic glasses with  maximal deep relaxation are  desirable. 

\section{Conclusions}

Parallel calorimetric studies and measurements of high-frequency shear modulus  of glassy  Zr$_{57}$Nb$_5$Al$_{10}$Cu$_{15.4}$Ni$_{12.6}$ (Vit106) and high-entropy Zr$_{35}$Hf$_{13}$Al$_{11}$Ag$_8$Ni$_8$Cu$_{25}$ are performed. The investigation was carried out on samples in the as-cast, moderately preannealed  and strongly preannealed states. In the latter case, preannealing was performed during 160 hr just below the glass transition and led to deep relaxation resulting in a strong peak of the excess heat capacity upon approaching the supercooled liquid state.

Considering the high-frequency shear modulus to be a major physical  parameter controlling the relaxation kinetics  and applying the Interstitialcy theory,  we  derived an expression for the excess heat capacity due to two interdependent processes, \textit{i})
a change in the concentration of interstitial-type defects (elastic dipoles), which alter the magnitude of the shear modulus and \textit{ii}) a change in the shear modulus, leading to a change in the energy barrier height for atomic rearrangements. The excess heat capacity can be then calculated using experimental data on the shear modulus, its first and second derivatives over temperature.  It is shown that the excess heat capacity thus determined  provides a very good description of experimental calorimetric excess heat capacity both below $T_g$ and upon approaching the glass transition for as-cast, moderately and deeply relaxed states of glass. 

It is argued that the heat capacity behaviors near the glass transition of as-cast and deeply relaxed MGs are qualitatively different. While the defects in the as-cast glass are frozen below the glass transition and unfreeze upon heating above it and  the corresponding defect concentrations are comparable, the deeply relaxed state is characterized by  a low defect concentration, which is strongly increased upon heating over the glass transition. We show that deep relaxation results in a sharp rise of the excess entropy upon approaching the glass transition. This fact implies that hypothetical infinitely deep relaxation of glass should result in a discontinuous jump-like increase of the excess entropy. This is a major feature of the phase transition of the first kind.  

The situation becomes similar to the melting of a crystal in terms of the Interstitialcy theory: in order for "melting" to occur, a large amount of mobile defects must be generated, which requires a corresponding amount of heat. It is concluded that the deeper the relaxation of a metallic glass, the closer the transition glass $\rightarrow$ supercooled liquid will be to the melting, i.e. to a phase transition of the first kind.

\section{Acknowledgments}
The work was supported by  Russian Science Foundation under the project No. 23-12-00162.

\end{document}